\documentstyle[pre,aps,epsf]{revtex}
\begin{document}
\title{Spiral Turbulence: From the Oxidation of CO on Pt(110) to 
Ventricular Fibrillation}
\author{Ashwin Pande$^1$, Sitabhra Sinha$^{1,2}$ and Rahul Pandit$^{1,2}$\\
$^1$Department of Physics, Indian Institute of Science,\\
Bangalore - 560012, India\\
$^2$Jawaharlal Nehru Centre for Advanced Scientific Research,\\ 
Bangalore - 560064, India.}
\maketitle
\abstract{We give a brief overview of systems that show spiral patterns 
and spatiotemporally chaotic states. We concentrate on two physical 
systems: (1) the oxidation of CO on Pt(110) and (2) ventricular 
fibrillation in hearts. The equations that have been suggested as simple 
models for these two different systems are closely related for they 
are both {\it excitable media}. We present these equations and give a 
short summary of the phenomena they yield.}
\maketitle

\section{Introduction}
\label{intro}
Spiral waves are found in a wide variety of physical, chemical, and 
biological systems. These include convection patterns in cylindrical 
cells \cite{cross}, patterns in the Belousov-Zhabotinsky chemical reactions 
\cite{cross,belou}, the oscillations of cyclic-AMP in aggregating 
{\it Dictyostelium Discoidium} amoebae \cite{cross,siegert}, the oxidation 
of carbon monoxide (CO) on surfaces of platinum single crystals \cite{hild}, 
calcium waves in the cell cytoplasm \cite{lechleiter} of {\it Xenophus} 
oocytes, and in ventricular fibrillation \cite{glass} which is perhaps the 
most dangerous form of cardiac arrhythmia. A variety of mathematical models 
have been used 
to describe these phenomena; we refer the reader to the review by Cross and 
Hohenberg for a recent discussion of many such models \cite{cross}. The 
models range from cellular automata \cite{tyson} to deterministic partial 
differential equations, such as the Complex Ginzburg Landau equation (for 
oscillatory chemical reactions near onset) and equations in the 
FitzHugh-Nagumo class (for CO oxidation and ventricular fibrillation). 
We concentrate on the FitzHugh-Nagumo types of equations here.
Our description is not very technical since the talk on which this article 
is based was aimed at an audience comprising scientists from different 
disciplines. We begin with a brief overview of phenomena followed by an 
introduction to equations in the FitzHugh-Nagumo class. We then present 
some results for the specific models suggested for the oxidation of 
CO on Pt(110) and ventricular fibrillation \cite{hild,panfilov} 
including our recent studies \cite{pune}. 

Many chemical reactions display spontaneous spatial self-organisation
if the reactors are not stirred and the reactants are fed at a 
constant rate.  Among these reactions the oxidation of CO on Pt(110) has been 
studied in great detail \cite{hild}. The experimental setup
consists of a cell containing a Pt wafer kept at a constant temperature.
The reactants are fed into the cell at a constant rate and the products
removed. Spatial structures in the concentration fields of the
reactants have been observed by using photoelectron emission 
microscopy (PEEM). For a picture of the growth of a spiral wave on 
such a Pt(110) wafer see Ref. \cite{hild}. 

A human heart consists of four chambers - the two upper atria and two lower 
ventricles. Heart activity is driven by periodic signals emitted by the cells 
of the sinus node situated in the upper right atrium. These cells initiate
the heartbeat cycle by emitting a pulse of electrical activity. This 
activity travels as a wave across the atria and stimulates the nerves
of the atrioventricular node (AVN). From the AVN the signal travels 
along the His bundle and into the ventricles via the Purkinje fibre
network which extensively innervates the ventricular wall. 
The ventricles now contract thus completing the beat cycle. 
Cardiac arrhythmia is caused when the signal from the sinus node does not
propagate in the normal manner. Especially dangerous is ventricular 
fibrillation, an irregular pulsation of the ventricles, which is a 
major cause of death in the industrialised world \cite{winchaos}. 
It has been conjectured for many years \cite{winf} that ventricular 
fibrillation is associated with the formation of spiral waves (or 
their three-dimensional analogues called scroll waves) on the 
walls of the ventricles. Experimental evidence for such spiral waves has 
been increasing with the advent of voltage-sensitive dyes and advanced 
imaging techniques \cite{nature}. Figure 2 shows such a spiral 
wave on the ventricle of a canine heart. Such spirals have also been 
seen to break up.
\begin{figure}
\epsfxsize=6.0in
\epsfysize=4.0in
\centerline{ \epsfbox{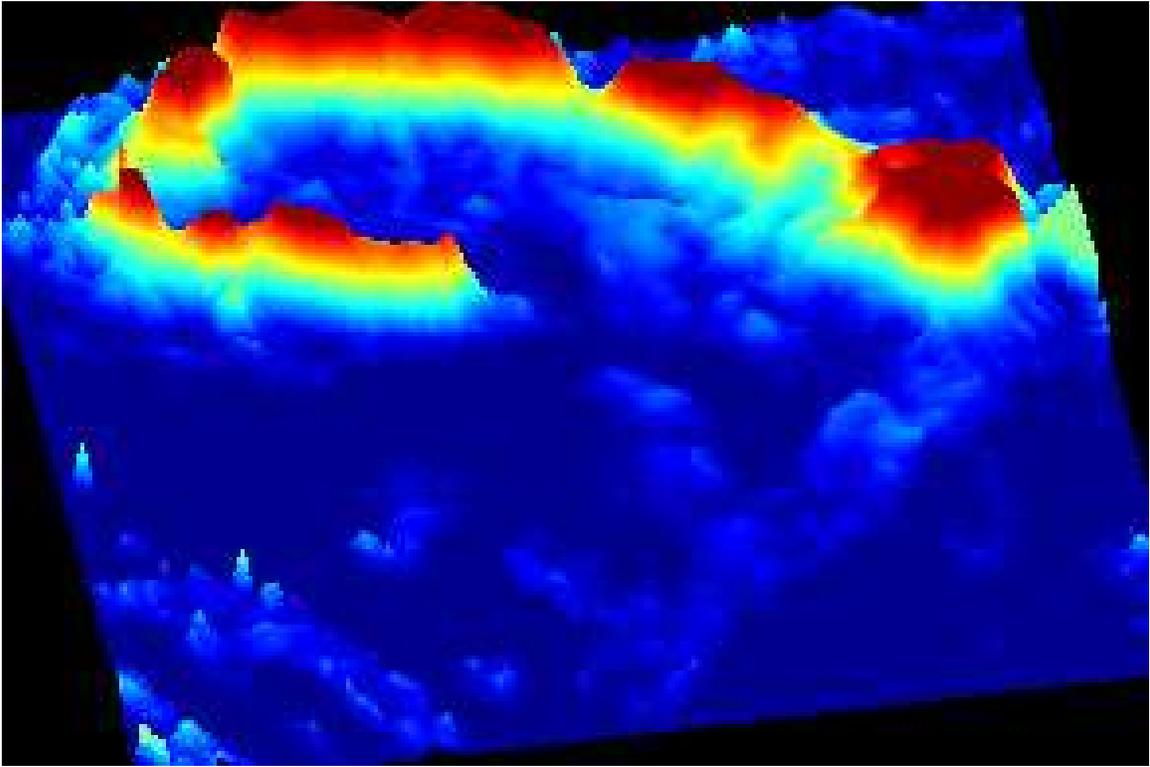} }
\caption{ A spiral wave on a canine heart imaged using voltage-sensitive dyes 
and CCD cameras, after Ref.{ \protect \cite{nature}} . 
Image obtained from and reproduced with the permission of W. Ditto. }
\end{figure}

   Both the oxidation of CO on Pt (110) and ventricular fibrillation 
have been the focus of many theoretical studies. These are  motivated 
partially by a desire to understand the phenomena seen in experiments 
and partially to elucidate the general properties of systems displaying
such spatiotemporally chaotic, statistical steady states. The remainder
of this paper is organised as follows: Section 2 consists of an introduction
to FitzHugh-Nagumo equations and the models we consider. 
Section 3 describes some results for the model of CO oxidation on Pt(110). 
Section 4 describes some results for the model of ventricular fibrillation. 
Section 5 ends with concluding remarks.
\section{Models}
\label{sec2}
  FitzHugh-Nagumo-type equations  belong to the general class
of reaction-diffusion equations. These are of the form
\begin{equation}
\frac{\partial U_i}{\partial t} = f_i({U}_j) + D_{ij} \nabla^2 U_j;
\end{equation}
where the $f_i$ are nonlinear functions of the $U_j$'s and $D_{ij}$ are
elements of a matrix of diffusion constants.
The FitzHugh-Nagumo-type system used by Hildebrand et. al. \cite{hild} 
for the oxidation of CO on Pt(110) is 
\begin{eqnarray}
\frac{\partial u}{\partial t} &=& \nabla^2 u -
\frac{1}{\varepsilon} u(u-1)(u - (v+b)/a) ,\nonumber \\
   \frac{\partial v}{\partial t} &=& f(u) - v ; \label{fhNag}
\end{eqnarray}
here the fields $u\/$ and $v\/$ are related to the CO coverage and
the surface reconstruction, $a, b,\/$ and $\varepsilon\/$ 
are control parameters.
 Physically, $\varepsilon$ is proportional to the ratio of the
rate constant for change in surface structure to the rate constant for
the oxidation of adsorbed CO on the surface.
Further,
  $ 
   f(u)  =  0  \mbox{ if } u < \frac{1}{3}\mbox{ , }
   f(u)  =  1 - 6.75u(u-1)^2 \mbox{ if } \frac{1}{3} \leq u < 1 \mbox{,  and }
   f(u)  = 1 \mbox{ if } u \geq 1.
  $ 
The form of $f(u)$ leads to the production of $v$ only above a 
certain threshold value of $u$.
This is an effective model with $u$ and $v$ related to the
reactant coverages and the surface reconstruction. We refer the reader to
Ref. \cite{hild} for further details and the specific form for $f(u)$,
which has been found to model the CO oxidation
on Pt(110) experiments well. We have studied Model (\ref{fhNag}) with both
periodic and Neumann (no-flux) boundary conditions. Here we restrict
ourselves to Neumann B.C.'s for ease of comparision with model (\ref{Panf}).

The Panfilov-Hogeweg model for ventricular fibrillation that we consider  
is \cite{panfilov} 
\begin{eqnarray}
\frac{\partial e}{\partial t} &=& \nabla^2 e - f(e) - g, \nonumber \\
\frac{\partial g}{\partial t} &=& \epsilon(e,g) ( k e - g); \label{Panf}
\end{eqnarray}
where $e$ is the transmembrane potential and $g$ is a recovery variable
corresponding to the amount of openness of the ion channels \cite{winf}. 
Also, $f(e) = C_1 e $ if $e < e_1$; $f(e) = -C_2 e + a$ if 
$e_1 \leq e \leq e_2$; $f(e) = C_3 (e-1)$ when $e > e_2$, and
$\epsilon(e,g) = \epsilon_1$ when $e < e_2$; $\epsilon(e,g) = \epsilon_2$
when $e > e_2$, and $\epsilon(e,g) = \epsilon_3$ when $e < e_1$ and $g < g_1$.
Physically $\epsilon(e,g)^{-1}$ determines the refractory time period
of this model of ventricular muscle.
Since the ventricles are electrically insulated from the atria, 
we impose Neumann boundary conditions. 
\section{CO oxidation on Pt(110)}
\label{sec3}
  
\begin{figure}[hb]
\epsfxsize=6.0in
\epsfysize=3.2in
\centerline{\epsfbox{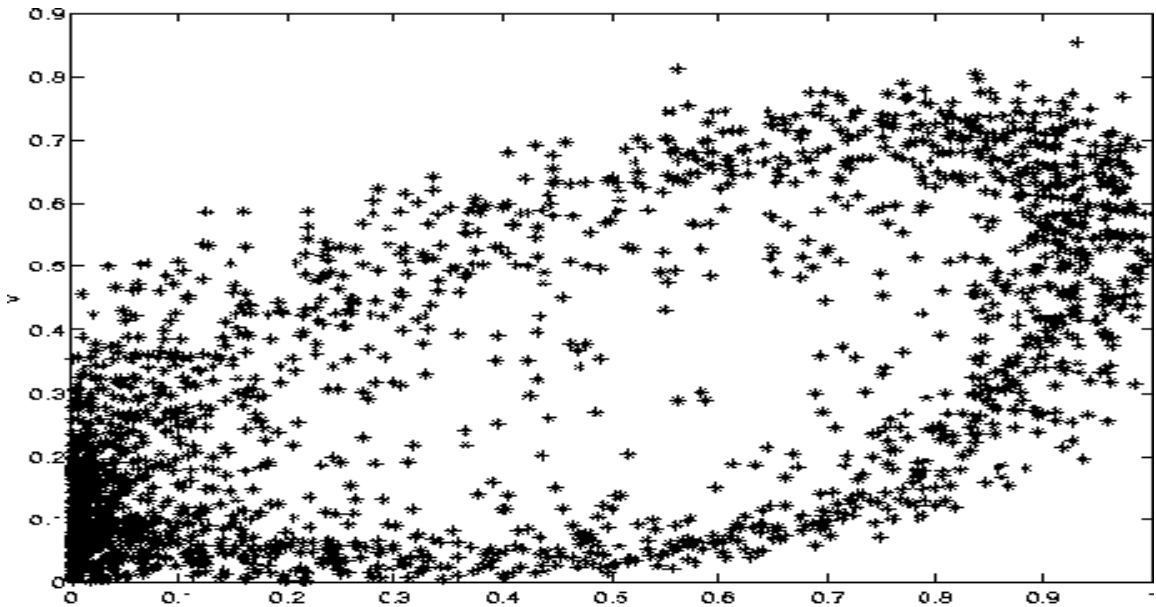}}
\caption{The local phase portrait in the state T1 of model (\ref{fhNag}) at 
$\varepsilon = 0.0781$. Such local phase portraits get progressively more 
disordered in the regimes M, T1 and T2.\label{lpp}}
\end{figure}

For the model of CO oxidation on Pt(110) (\ref{fhNag}), Hildebrand et. al.
\cite{hild}, find that there are many different statistical steady states.
These include states with no waves (N), states with flat shrinking waves (F),
states with rigidly rotating spirals (S), states with meandering 
spirals (M) and turbulent states with creation and annihilation of
spirals (T1 and T2). These authors have found a stability diagram 
or "phase diagram" in the $b-\varepsilon$ plane for $a = 0.84$, 
and have calculated the spatial autocorrelation function
of the density of spiral cores 
$\langle n(\vec{x},t) n(\vec{x} + \vec{s}, t + \tau)\rangle \/$.
They suggest that the change in the spiral dynamics on passing from T1 to
T2 is like a "liquid-gas" transition. 
We have simulated model (\ref{fhNag}) by discretising it on a square
lattice and using a variable stepsize fourth-fifth order Runge-Kutta
integrator for time stepping \cite{numrec}. In some cases we also use a
fast integration scheme proposed by Barkley \cite{bark}. We have used 
Neumann boundary conditions to compare with model (\ref{Panf}).
We have investigated the local dynamics, i.e., the dynamics obtained
by plotting $u(\vec{x_n},t_n)$ versus $v(\vec{x_n},t_n), n= 1,2,\ldots$
for fixed spatial location $\vec{x_n}$ and $t_n= n \Delta$ where $\Delta$ is 
a sampling interval which we choose to be $=0.1$ time units. Such local
phase portraits show that, in the states S, M, and T1, the system displays
oscillatory behaviour; a representative phase portrait is shown in 
Fig. (\ref{lpp}) for T1.
The trajectory in the local phase portrait loops around the fixed point
$(u_\ast,v_\ast)\equiv(0.66, 0.484)$
of equation (\ref{fhNag}) without the $\nabla^2$ term. 
We define the phase $ \phi({\bf{x}},t) \equiv 
\tan^{-1}((v({\bf{x}},t) - v_\ast) / (u({\bf{x}},t) - u_\ast))\/$
and note that it winds by $2\pi\/$ around the cores of spiral defects 
\cite{hild}. Thus it can be used to obtain the defect density
$\rho$ \cite{pune}.
%.................
\begin{figure}[ht]
\epsfxsize=6.0in
\epsfysize=3.0in
\centerline{\epsfbox{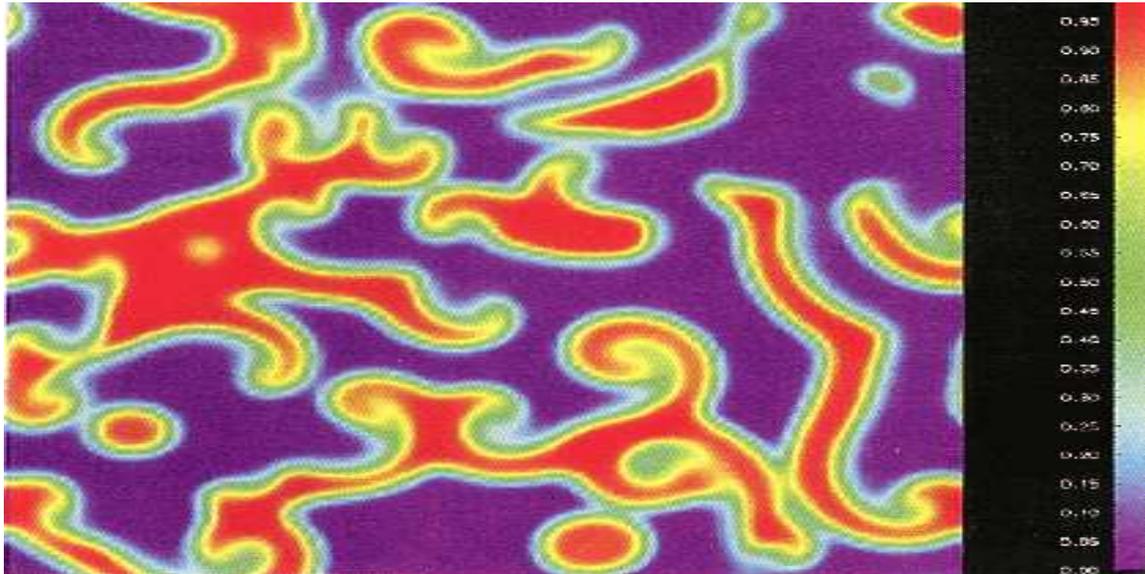}}
\caption{A plot of $u(\vec{x})$ as a function of space in model (\ref{fhNag}).
The colour at a point is proportional to the value of $u\/$ at that
point, obtained after evolving a typical initial condition for 5000 
time units \label{uT1plot}.
Note the similarity of this figure with Fig (\ref{peplot}). For this
figure $\varepsilon = 0.073$ which corresponds to the T1 state.}
\end{figure}
%...............
\begin{figure}[ht]
\epsfxsize=6.0in
\epsfysize=3.0in
\centerline{\epsfbox{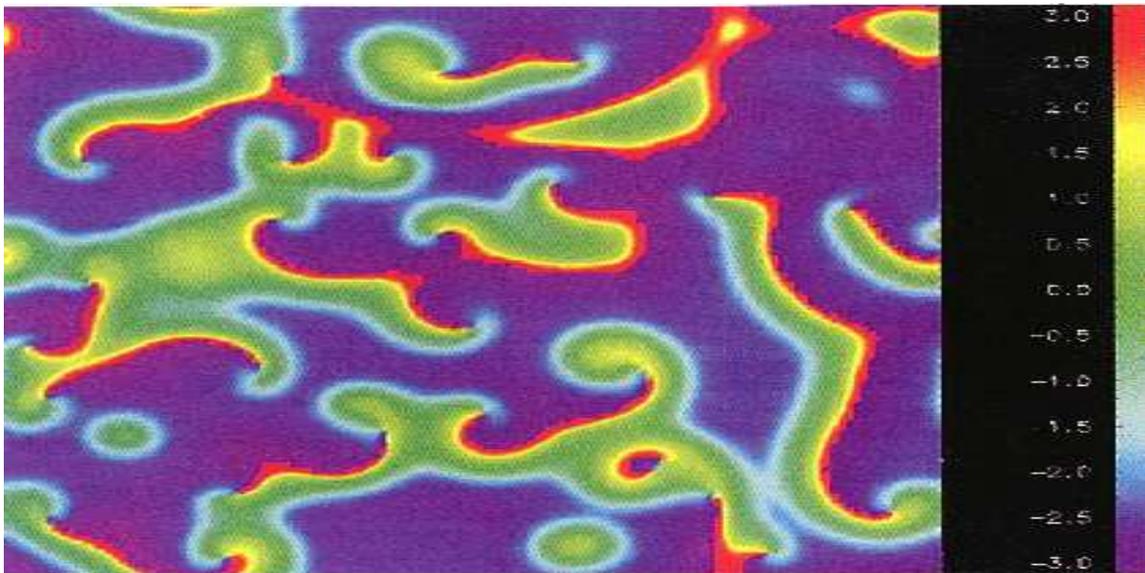}}
\caption{The phase $\phi\/$ (see text) corresponding to the field configuration
of Fig. (\ref{uT1plot}) as a function of space. The colour at a point
is proportional to the value of the phase at that point. \label{phT1plot} }
\end{figure}
%.............
\section{Ventricular Fibrillation}
\label{sec4}
\begin{figure}[hb]
\epsfxsize=6.0in
\epsfysize=4.0in
\centerline{\epsfbox{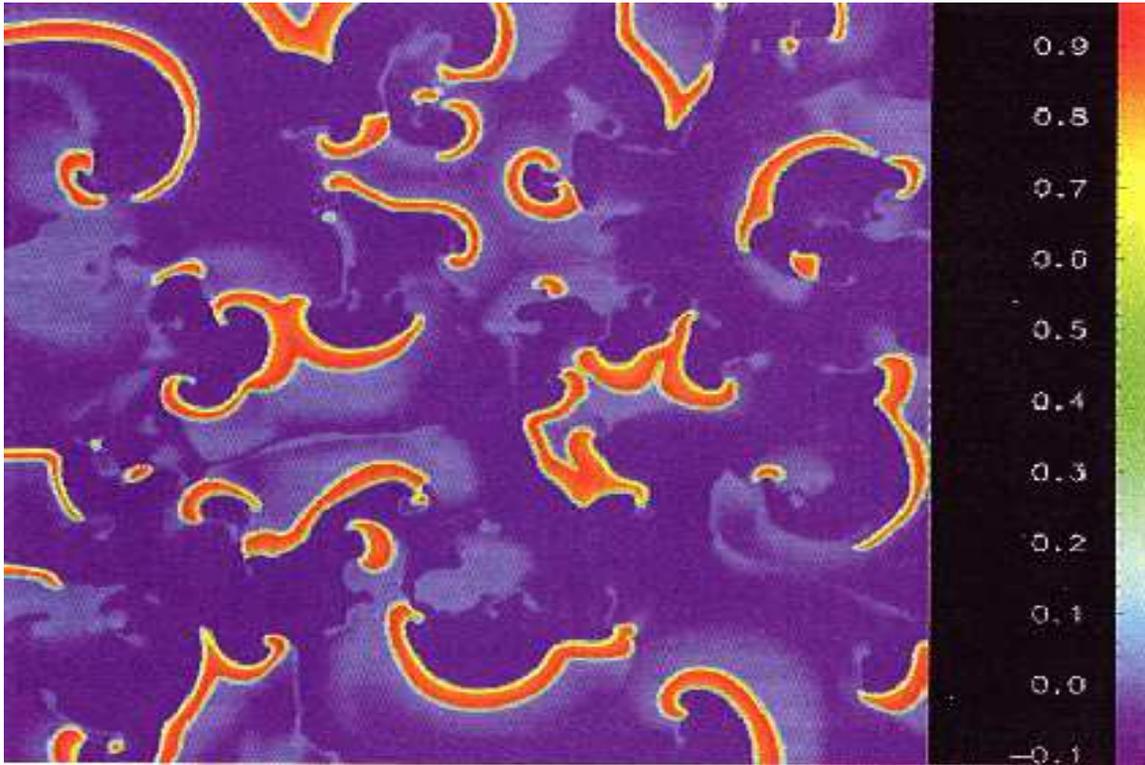}}
\caption{The configuration of the $e\/$ field of model (\ref{Panf})
on a system of size $256 \times 256$ obtained after evolving a typical
initial condition for 1000 time units. \label{peplot}}
\end{figure}

\begin{figure}[ht]
\epsfxsize=6.0in
\epsfysize=4.0in
\centerline{\epsfbox{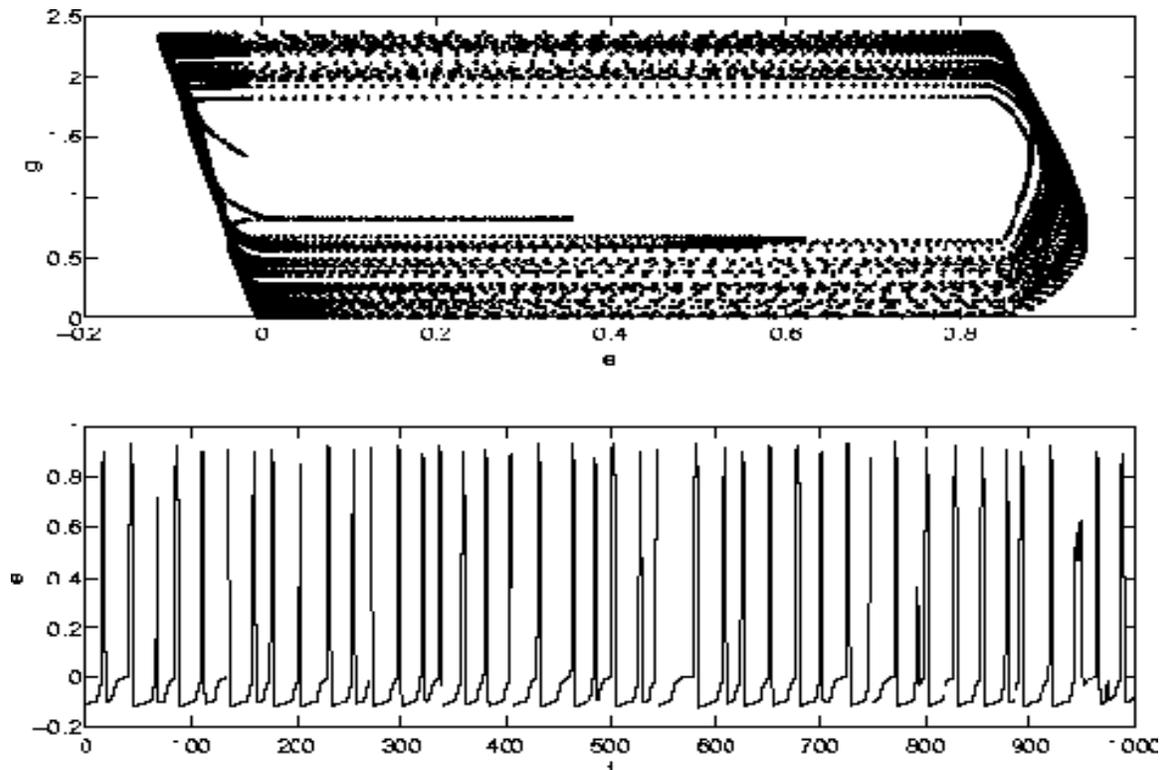}}
\caption{ The upper figure shows a local phase portrait of the 
model (\ref{Panf}); the lower shows a time series of the transmembrane  
potential $e\/$. The system size was $256 \times 256$ with
a grid spacing of $0.5$. \label{plpPlt} }
\end{figure}

It is believed that ventricular fibrillation occurs because of 
the spontaneous breakdown of contraction waves in the heart muscle leading to
a state with many rotating spirals \cite{winf,nature}.
Physiological abnormalities do not seem to play an important role
in this arrhythmia in a sizeable fraction of cases. 
Thus it is interesting to study models in which spiral breakup
appears {\em spontaneously} without introducing spatial inhomogenieties
in the diffusion constant, for example. 
For the model (\ref{Panf}), Panfilov et. al.
\cite{panfilov} find that, for the parameter values
$e_1 = 0.0026, e_2 = 0.837, C_1 = 20, C_2 = 3, C_3 = 15, a = 0.06,
k=3,{\epsilon_1}^{-1} = 75, {\epsilon_2}^{-1} = 1, g_1 = 1.8 $ 
and $0.5 < {\epsilon_3}^{-1} < 10$, an initial condition in the shape of
a broken wave gives rise to states containing many spirals.
They have also calculated an electrocardiogram
(ECG) numerically for this model in the turbulent state. 
In this section we elucidate the spatiotemporal nature of the chaotic
behaviour of this system. In Ref. \cite{panfilov} it has been noted
that the breakdown of spirals in models (\ref{fhNag}) and (\ref{Panf}) 
is qualitatively similar. We elaborate on this similarity below.
The state with broken spirals in model (\ref{Panf}) is similar to the state T1
in model (\ref{fhNag}) as can be seen by comparing Figs. (\ref{peplot}) and
(\ref{uT1plot}). 
We have simulated model (\ref{Panf}) using finite-differencing in space
with a forward euler algorithm for time stepping. Most of our
simulations are done with a system size of $256 \times 256$ with 
a spatial grid step of $dx=0.5$ and a time step of $dt=0.022$.
Physically, one time unit is equal to 5 milliseconds, while one 
spatial unit is equal to 1 millimeter. We use the parameter values given 
above. We have calculated a local phase portrait for this model, i.e.
a plot of $e(\vec{x}_n,t_n)$ vs $g(\vec{x}_n,t_n)$ for a fixed spatial 
location $\vec{x}_n$. This is shown in Fig. (\ref{plpPlt}). Note that this 
local phase portrait is similar to that in the T1 state of the CO 
oxidation model (\ref{fhNag}).
We find that for a system of size $50 \times 50$ the lifetime $\tau_t$ 
of the turbulent state is $\sim 300$ time units. After this time the entire 
field configuration goes to zero for various initial conditions. 
For a system of size $100 \times 100$ $\tau_t \sim 950 $ time units.
We conjecture that $\tau_t$ grows with system size; we are investigating
this scaling of $\tau_t$. The typical size of a human heart
corresponds to a system size of $\sim 600 \times 600$  so the size dependence
of $\tau_t$ is of obvious interest. 
We note in passing that this qualitative trend of $\tau_t$ increasing with
system size is in accord with the well-known phenomenon that only the
larger mammals have heart attacks.

\begin{figure}[hb]
\epsfxsize=6.0in
\epsfysize=3.0in
\centerline{\epsfbox{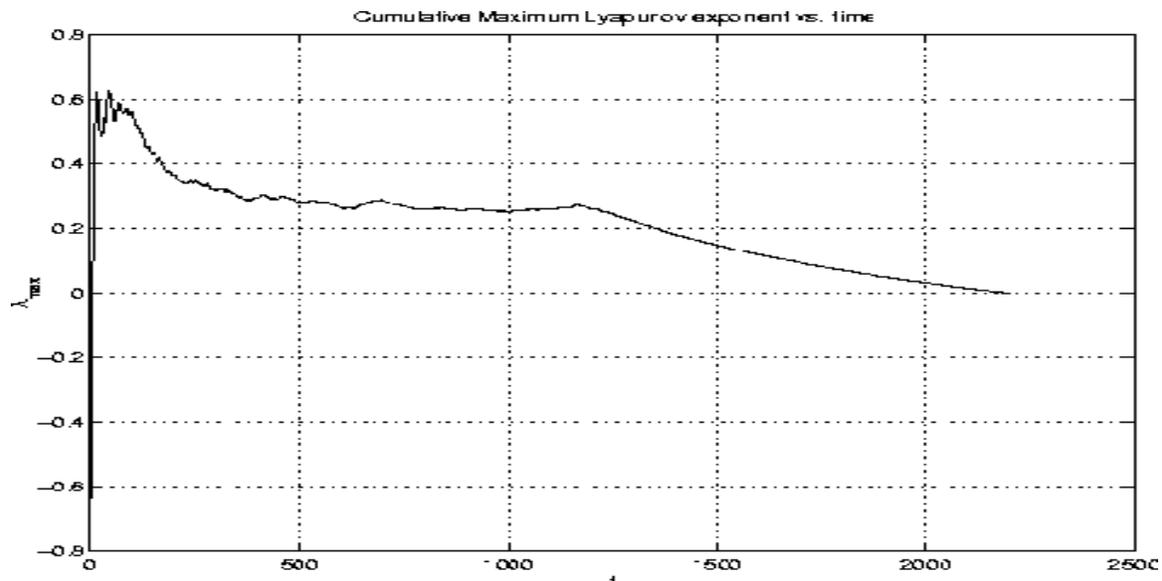}}
\caption{ The approximation to the maximum Lyapunov exponent $\lambda_{max}\/$ 
at time $t\/$ versus $t\/$. Note that $\lambda_{max}\/$ approaches a 
positive constant ($\simeq 0.2\/$) and then decays at large times to negative 
values. This occurs because there is a long-lived chaotic transient which 
finally decays to a state that is constant in space with $e(\vec{x},t) = 0$ and
$g(\vec{x},t) = 0$ everywhere. The lifetime of this chaotic transient 
increases with the size of the system (see text).\label{pLyap}}
\end{figure}

To characterise chaos in model (\ref{Panf}), we have calculated the 
maximum Lyapunov exponent $\lambda_{max}$ for a system of size
$128\times 128$.
The maximum Lyapunov exponent measures the divergence of nearby
trajectories in the system. To calculate it we use standard 
methods \cite{chua}. 
In the transient turbulent state of model (\ref{Panf}), $\lambda_{max}$
saturates to a value of $\sim 0.2$. This is shown in Fig (\ref{pLyap}). Note
that when this transient decays, the maximum Lyapunov Exponent decreases
to negative values. 

\section{Conclusions}
\label{sec5}
Spiral patterns are seen in a wide variety of natural systems \cite{cross}. 
Here we have concentrated on two such systems which are also excitable media. 
The chaotic states of these systems display the breakup of large spirals 
into smaller ones in the states T1 and T2. The qualitative similarity of 
such spiral breakup in models (\ref{fhNag}) and (\ref{Panf}) has been mentioned 
briefly earlier \cite{panfilov}. Here we have elucidated this similarity by 
comparing the spiral patterns (Figs. \ref{uT1plot} and \ref{peplot})
in the state T1 of model (\ref{fhNag}) and the chaotic state of
model (\ref{Panf}); local phase portraits (Figs. \ref{lpp} and \ref{plpPlt})
are also similar.
We have characterised the chaos in model (\ref{Panf}) by computing the maximum
Lyapunov exponent $\lambda_{max}\/$ (Fig. \ref{pLyap}) as we have done for
model (\ref{fhNag}) elsewhere \cite{lopap}.  
 
Our study shows that there is considerable similarity between the behaviour 
of model (\ref{fhNag}) in the state T1 and model (\ref{Panf}) for the parameter 
values given above.  It should be possible to exploit this similarity to 
gain new insights about these two excitable media (e.g., by doing 
experiments on the oxidation of CO on Pt(110) which might be hard to 
do on a human heart). Of course some caution must be exercised while 
doing this for the two models are different in some respects: the chaotic 
state of model (\ref{Panf}) is a transient for the system sizes we have 
studied, whereas the analogous state for model (\ref{fhNag}) seems to be a 
statistical steady state; furthermore, if the $\nabla^2\/$ terms are dropped 
in both models, then model (\ref{fhNag}) has an extra unstable fixed point 
[at $(u,v) = (u_*,v_*)\/$], i.e., the reaction kinetics are different in the  
two models. We hope our work will stimulate experimental studies that 
will try to elucidate the similarities and differences between spiral 
turbulence in the oxidation of CO on Pt(110) and ventricular fibrillation. 
Since ventricular fibrillation in cardiac muscle is
fatal \cite{winf,glass,winchaos}, any insight gained about it by studying 
similar systems is of great interest. 

We thank CSIR  and JNCASR for support and SERC (IISc) for computational 
facilities.

\end{document}